# Novel high symmetry super-hard $C_{48}$ and $C_{32}$ allotropes with "**ana**" and "**ukc**" original topologies: Crystal chemistry and DFT investigations.

Samir F. Matar

Lebanese German University (LGU), Sahel Alma, P. O. Box 206 Jounieh, Lebanon

https://orcid.org/0000-0001-5419-358X

**Abstract.**

*Novel high symmetry body centered carbon allotropes: cubic $C_{48}$ and tetragonal $C_{32}$ are proposed with respective original **ana** and **ukc** topologies. Devised from crystal structure engineering, their ground state structures and energy derived physical properties were accurately derived based on quantum mechanics calculations within the density functional theory DFT. Both allotropes made of distorted tetrahedral C4 arrangements were found dense with $\rho > 3$ g/cm³ that remain lower than diamond's ($\rho = 3.50$ g/cm³). With cohesive albeit with metastable ground state structures, both allotropes show stability from the mechanical (elastic properties) and dynamic (phonons band structures) properties. Vickers hardness magnitudes $H_V(C_{48}) = 47$ GPa and $H_V(C_{32}) = 59$ GPa point to super-hard materials. The electronic band structures range from large direct band gap ~5 eV for $C_{48}$ to indirect band gap ~2.5 eV semi-conducting $C_{32}$. Such findings of original allotropes with targeted physical properties are bound to enrich the field of research on carbon.*

**Introduction**

Searching for carbon allotropes remains an active research field among scientists, with a focus on the systems approaching diamond's physical properties, like the mechanical and the electronic ones. Despite their mainly theoretical investigations, the works target applications *in fine*. Many allotropes are found "artificially" thanks to machined learned programs such as CALYPSO [1] and USPEX [2]. Nevertheless, approaches based on crystal structure engineering also help find allotropes with relevant properties as the transformation mechanisms 2D –soft- ➔3D –hard-, (D designating dimension) [3] or those qualified as super-hard due original small building units as *cyclopropanoïd C3* [4].

The web-accessible SACADA database [5] serves as a library of carbon structures, categorized in different topologies using TopCryst program [6]. For instance, diamond (cubic) is labeled "**dia**", and its rare hexagonal form occurrence "*Lonsdaleite*", is labeled "**lon**". Diamond known as the hardest material with Vickers hardness amounting to $H_V$~95 GPa is characterized by a high density ρ~3.55 g/cm$^3$ arising from the covalent character of the C-C short connections ($d_{C-C}$ ~ 1.54 Å) within $C(sp^3)$-like perfectly tetrahedral carbon. Nevertheless, an exceptional feature was observed for artificial carbon allotrope **qtz** $C_6$ (i.e., quartz-based topology) [7] found denser than diamond, with ρ = 3.67 g/cm$^3$ and subsequent larger Vickers hardness: $H_V$~100 GPa; it is now catalogued in SACADA data base under No. 1190 (https://www.sacada.info/ca_data.php?id=1190 ). With respect to diamond, **qtz** $C_6$, like other carbon artificial allotropes, is *metastable* being less cohesive; this is also shown herein for the two novel allotropes. Another feature can be observed in tetrahedral allotropes, namely the distortion of the *C4* tetrahedron with angles departing from the perfect ∠C–C–C angle of 109.47$^o$. This is also observed for presently devised and investigated allotropes. In this work, starting from crystal structure engineering of tetrahedral stacking, we identify high symmetry body-centered super-hard $C_{48}$ and $C_{32}$ with "**ana**" and "**ukc**" original topologies not documented in databases. The physical properties were derived from quantum mechanics calculations reporting on their cohesiveness, electronic, mechanical, and dynamic properties. After this Introduction, the paper is organized as follows: the Theoretical framework and the Computational methodology are given in Section 1; the Crystal Structure characteristics are presented in Section 2; the Mechanical properties from the elastic constants are addressed in Section 3; The Dynamic properties from the Phonons are detailed in Section 4. Section 5 presents the electronic band structures. The paper is ended with a Conclusion.

1- **Theoretical framework and Computational methodology**

To determine the ground state structures of the devised allotropes corresponding to energy minima and to derive the mechanical, dynamic properties and the electronic structures, quantum mechanics computations were carried out in the framework of the density functional theory DFT [8, 9]. Among calculation methods within the DFT, use was made of the Vienna Ab initio Simulation Package (VASP) code [10, 11] with atomic potentials from the Projector Augmented Wave (PAW) method [11, 12]. DFT exchange-correlation (XC) effects were considered within the generalized gradient approximation (GGA) [13]. Relaxation of the atoms to the ground state structures was performed with the conjugate gradient algorithm according to Press *et al*. [14]. The Bloechl tetrahedron method [15] with corrections according to the scheme of Methfessel and Paxton [16] was used for geometry optimization and energy calculations. Brillouin-zone (BZ) integrals were approximated by a special **k**-point sampling according to Monkhorst and Pack [17]. Structural parameters were optimized until atomic forces were below 0.02 eV/Å and all stress components $< 0.003$ eV/Å$^3$. The calculations were converged at an energy cutoff of 400 eV for the plane-wave basis set in terms of the automatic high precision **k**-point integration in the reciprocal space to obtain a final convergence and relaxation to zero strains for the original stoichiometries presented in this work. In the post-processing of the ground state electronic structures, the charge density projections were operated on the lattice sites.

The mechanical stability was obtained from the elastic constants (Cij) calculations. The treatment of Cij results to extract the mechanical properties was operated thanks to the ELATE online program [18]. The outcome provides the bulk (B) and shear (G) modules along different averaging methods; the Voigt method [19] was used here for $B_V$ and $G_V$. The methods of microscopic theory of hardness by Tian et al. [20] and Chen et al. [21] were used to estimate the Vickers hardness ($H_V$) from the bulk and shear modules $B_V$ and $G_V$ (*vide infra*).

The dynamic stabilities were confirmed by calculating the phonons band structures all presenting positive phonon frequencies. The corresponding phonon band structures were obtained from a high resolution of the orthorhombic Brillouin zone according to Togo *et al*. [22]. The electronic band structures were obtained using the DFT-based augmented spherical wave ASW method [23] with GGA approximation for the treatment of the XC effects [13]. The VESTA (Visualization for Electronic and Structural Analysis) program [24] was used to visualize the crystal structures.

## 2- Crystal chemistry

Original $C_{48}$ was characterized in high symmetry body center cubic (BCC) space group *Ia*-3 No. 230. Running Topcryst [6] program led to **ana** topology, undocumented in SACADA or any other database to the best of our knowledge. The lattice parameters are given in Table 1 and the structure in ball-and-stick and polyhedral projections are given in Fig. 1a. The latter projection exhibits distorted tetrahedra with angles departing from 109.47° as shown in Table 1, but the shortest C-C separation is 1.54 Å, close to diamond interatomic value. This can be expected is so far that the coordination polyhedra are all tetrahedra. All 48 carbon atoms belong to the 48*g* Wyckoff position. The cohesive energy of -1.76 eV/atom is lower than diamond's which amounts to -2.47 eV/atom. Then $C_{48}$ is a metastable system *versus* diamond. The calculated density of 3.17 g/cm$^3$ remains lower than diamond's.

The second allotrope $C_{32}$ was identified in body center tetragonal (BCT) space group *I*4$_1$/acd (No. 142). Following the same protocol as for $C_{48}$, for identifying the topology, **ukc** label was found. **ukc** is also not identified in SACADA or another database, whence the originality of the work. The structure is shown in Figure 1b, and Table 1 presents the ground state structural parameters. $C_{32}$ is characterized with a higher density of ρ=3.34 g/cm$^3$ due to shorter d(C-C) =1.52 Å and a smaller atomic volume versus $C_{48}$. The building blocks are distorted *C4* tetrahedra shown in Fig. 1b with the polyhedral representation. The angles amount to ∠C–C–C = 89° and 103°. As in the first allotrope, all atoms belong to a single Wyckoff position: 32*g*. The cohesive energy is better but remains below diamond's and $C_{32}$ is also a metastable allotrope. Yet, it will be shown in the next sections that both allotropes are stable mechanically and dynamically whilst presenting interesting electronic structures related to diamond.

*Charge density description*.

The tetrahedral character of the carbon environment is better assessed from the projections of the charge density volumes. Figure 2 shows with yellow volumes the charge density around and between the carbon atoms. The tetrahedral distorted shapes of the tetrahedra can be observed in the two subfigures. Furthermore, the charge densities exhibit the characteristics of being separated from each other. Such localization will be further illustrated in the electronic band structures in Section 5.

## 3- Mechanical properties

The investigation of the mechanical properties was carried out by performing finite distortions of the lattice and deriving the elastic constants from the strain-stress relationship. The calculated sets of elastic constants $C_{ij}$ (i and j indicating directions) are given in Table 2. All $C_{ij}$ values are positive signaling stability of the two allotropes. Using ELATE program [18], the bulk B and the shear G modules obtained by averaging the elastic constants using different methods; Voigt's [19] are given herein, namely $B_V$ and $G_V$.

$B_V$ values show higher magnitudes than $G_V$. Larger magnitudes of both are observed for $C_{32}$ *versus* $C_{48}$. Nevertheless, these values remain lower than diamond's: $B_V$ =444 GPa and $G_V$= 534 GPa [25].

The corresponding Pugh ratios $G_V/B_V$ [26] relevant to "ductile-to-brittle" criterion were then calculated. For $G_V/B_V < 1$ a trend to ductile behavior is expected whereas for $G_V/B_V > 1$, a brittle behavior is expected. Indeed, for diamond, known for its high brittleness, $G_V/B_V$ = 1.20. The Pugh ratios for the two carbon allotropes are below unity letting expect moderate ductility. Nevertheless, a value close to 1 for $C_{32}$ indicates a tendency to brittleness. Such trends are translated into differences of Vickers hardness Hv calculated along with two models of microscopic theory of hardness:

$Hv^1 = 0.92(G_V/B_V)^{1.137} G_V^{0.708}$      (Tian et al.) [20]

$H_V^2 = 2(G_V^3/B_V^2)^{0.585} - 3$      (Chen et al.) [21]

The corresponding Vickers hardness ($H_V$) magnitudes obtained along the two methods are given in the last two columns of Table 2. They show close magnitudes with a larger hardness characterizing $C_{32}$. But both allotropes can be considered a super-hard.

## 4- Dynamic and thermodynamic properties

The dynamic stability of the two carbon allotropes was verified with the analysis of their phonon properties. The phonon band structures obtained from a high resolution of the Brillouin zone (BZ) in accordance with the method proposed by Togo *et al*. [22] are shown in Figure 3. The bands (red lines) develop along the main directions of the respective BZ, i.e. cubic and tetragonal, along the horizontal *x*-axis. Vertical lines separate them for better visualization. The vertical direction (*y*-axis) represents the frequencies ω, given in terahertz

(THz). Due to the large number of atoms in the two allotropes, body center cells were considered in the calculations.

For each crystal system the phonons band structures include 3N bands (N number of atoms) describing three acoustic modes starting from zero energy ($\omega = 0$) at the $\Gamma$ point (the center of the Brillouin zone) and reaching up to a few terahertz, and 3N-3 optical modes at higher energies. The low-frequency acoustic modes are associated with the rigid translation modes (two transverse and one longitudinal) of the crystal lattice. Most instability in systems arises from negative low frequencies that are not observed here. The calculated phonon frequencies are all positive, indicating that the allotropes are dynamically stable. The highest bands are observed around ~40 THz. This magnitude is close to the value observed for diamond by Raman spectroscopy [27], letting expect some relationship with diamond, as for the electronic structure as shown below.

### 5- Electronic band structures

Using the crystal parameters in Table 1, the electronic band structures were obtained for the two carbon allotropes using the all-electrons DFT-based augmented spherical wave method (ASW) [23] and GGA XC approximation [13]. Here too, body centered cells (half number of atoms) were considered in the calculations.

The band structures are displayed in Figure 4. The bands develop along the main directions of the cubic and tetragonal Brillouin zones. Along the vertical direction both allotropes present a band gap separating the filled valence band (VB) from the empty conductions band (CB), so that the energy zero is with respect to the top of VB, $E_V$. Fig. 4a exhibits an energy gap signaling a largely insulating character for $C_{48}$ with a gap larger than 5 eV and characterized by a direct character between $\Gamma_V$ and $\Gamma_C$. Oppositely, a semi-conducting behavior is observed for $C_{32}$ with an indirect gap between $Z_V$ and $\Gamma_C$. Such electronic band structures exhibiting band gaps go along with the particular features discussed with the analyses of the change densities at the end of Section 2.

**Conclusion**

Based on crystal chemistry rationale and quantum mechanics calculations of ground state crystal structures and pertaining physical properties, two high stoichiometry body centered cubic and tetragonal $C_{48}$ and $C_{32}$ were proposed. Specifically, the structures were identified with distorted *C4* tetrahedra and high density leading to super-hard behaviors derived from their stable elastic properties. Dynamically, both allotropes were found stable with positive frequencies revealed from their phonons band structures with highest magnitudes around 40 THz, a frequency characterizing diamond experimental Raman results. The electronic band structures reveal largely insulating $C_{48}$ and semi-conducting $C_{32}$.

**Tables**

Table 1. Crystal structure properties of novel cubic $C_{48}$ and tetragonal $C_{32}$

| Topology Allotrope Space Group No. | **ana*** $C_{48}$ cubic *Ia*-3 No. 230 | **ukc**** $C_{32}$ tetragonal *I*41/*acd* No. 142 |
|---|---|---|
| a / c, Å | 6.7087 | 4.6946 / 8.7048 |
| ∠C–C–C° | 85/125 | 89/103 |
| Shortest dist. Å | 1.54 | 1.52 |
| Volume, Å$^3$ | 301.93 | 191.85 |
| V/at., Å$^3$ | 6.29 | 5.98 |
| Density ρ (g/cm$^3$) | 3.17 | 3.34 |
| Atomic positions | C(48*g*) 0.6588, 0.0912, 0.875 | C(32*g*) 0.6283, 0.0979, 0.9094 |
| $E_{total}$, eV | -401.27 | -271.84 |
| $E_{tot.}$/at., eV | -8.36 | -8.49 |
| $E_{coh}$/at., eV | -1.76 | -1.89 |

***ana**;4/4/c5; sqc11219; ANA

****ukc**; 4/4/t73

Table 2. Elastic Constants and Voigt-averaged properties of bulk $B_V$ and shear $G_V$ modules and Vickers hardness $H_V$ calculated along two models. All quantities are expressed in GPa units (cf. text).

| $C_{ij}$ | $C_{11}$ | $C_{22}$ | $C_{12}$ | $C_{13}$ | $C_{33}$ | $C_{44}$ | $C_{55}$ | $C_{66}$ | $B_V$ | $G_V$ | $G_V/B_V$ | $H_V^1$ | $H_V^2$ |
|---|---|---|---|---|---|---|---|---|---|---|---|---|---|
| $C_{48}$ | 710 | 710 | 149 | 149 | 710 | 311 | 311 | 311 | 336 | 299 | 0.89 | 46 | 47 |
| $C_{32}$ | 965 | 972 | 91 | 129 | 759 | 399 | 313 | 331 | 376 | 365 | 0.97 | 58 | 59 |

FIGURES

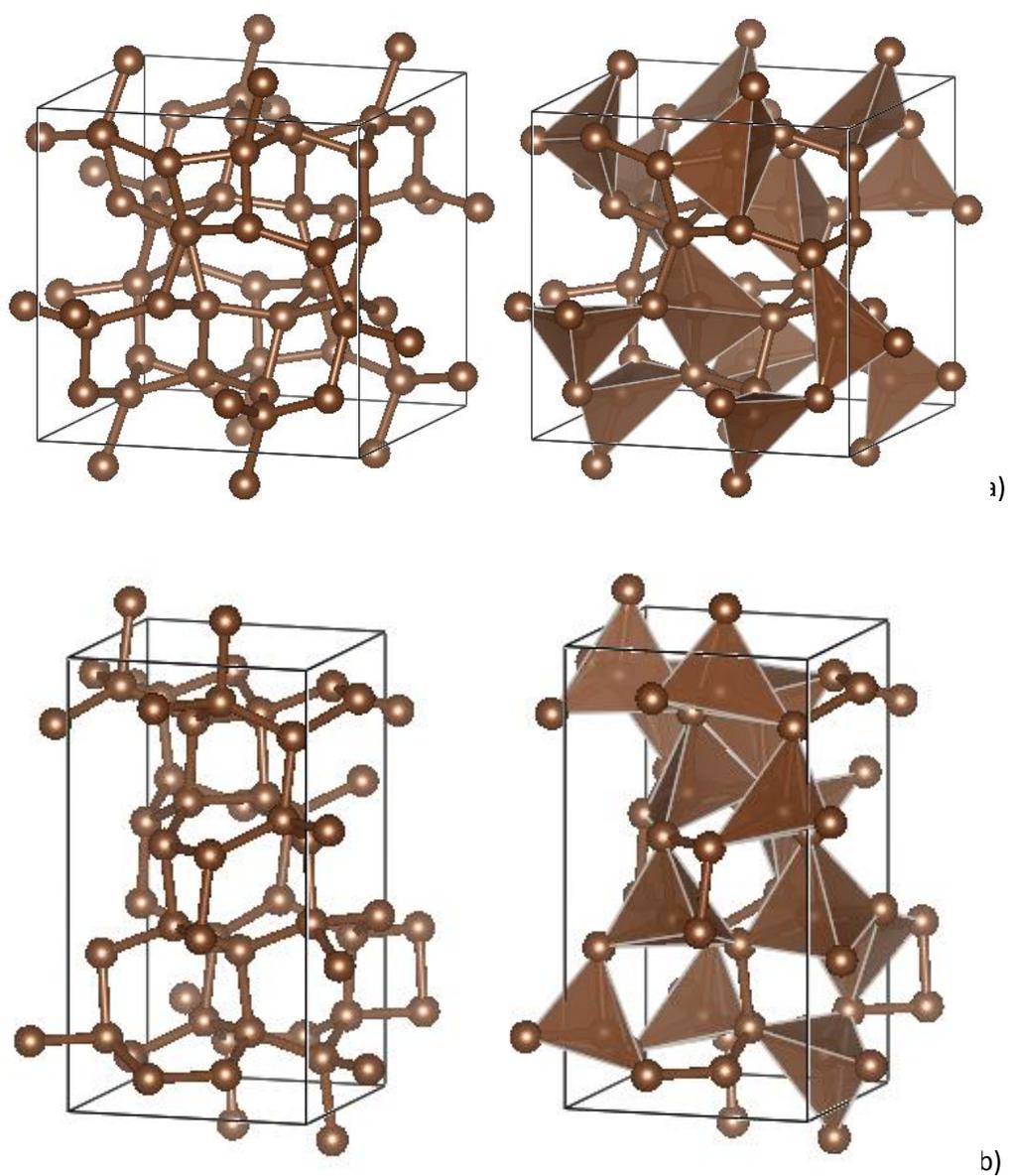

Figure 1. Crystal structures in ball-and-stick (left) and polyhedral (right) representations: a) $C_{48}$, b) $C_{32}$.

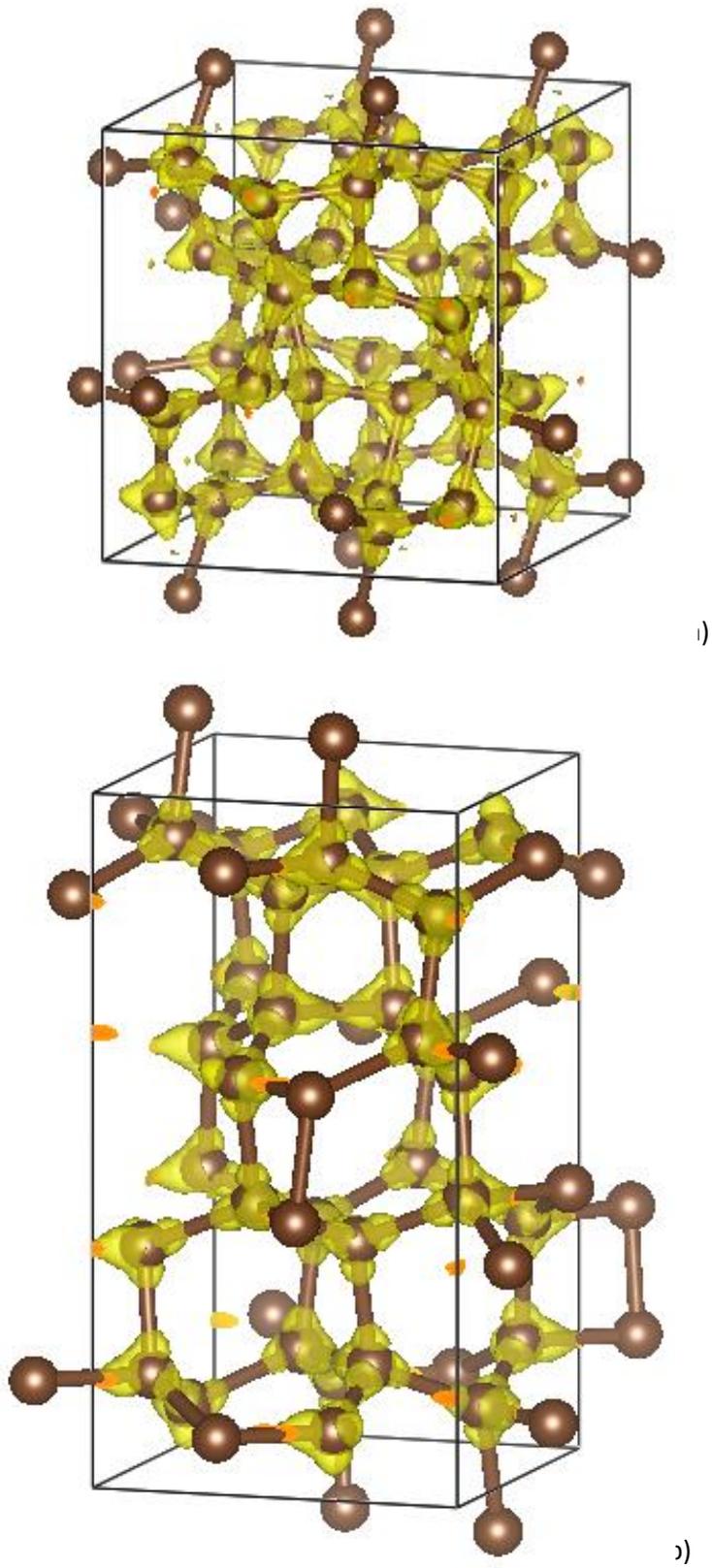

Figure 2. Charge density projections (yellow volumes) exhibiting tetrahedral configuration around carbon atoms: a) $C_{48}$, b) $C_{32}$.

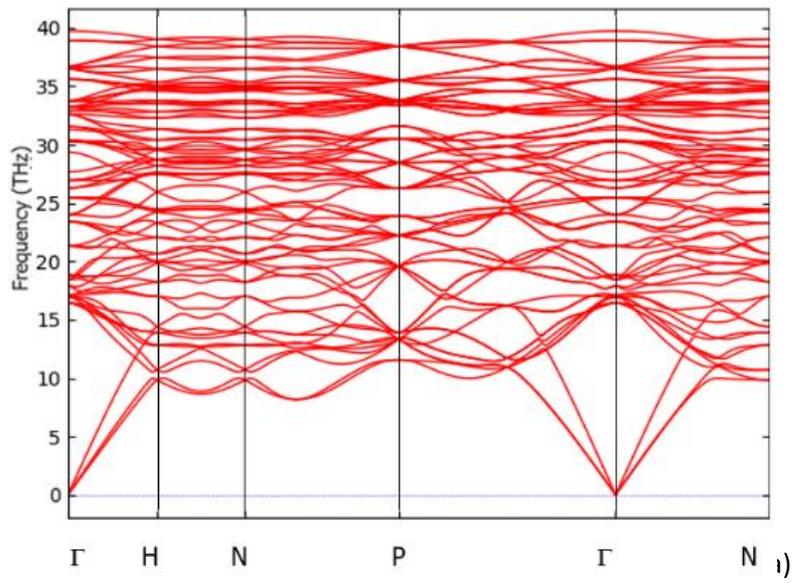

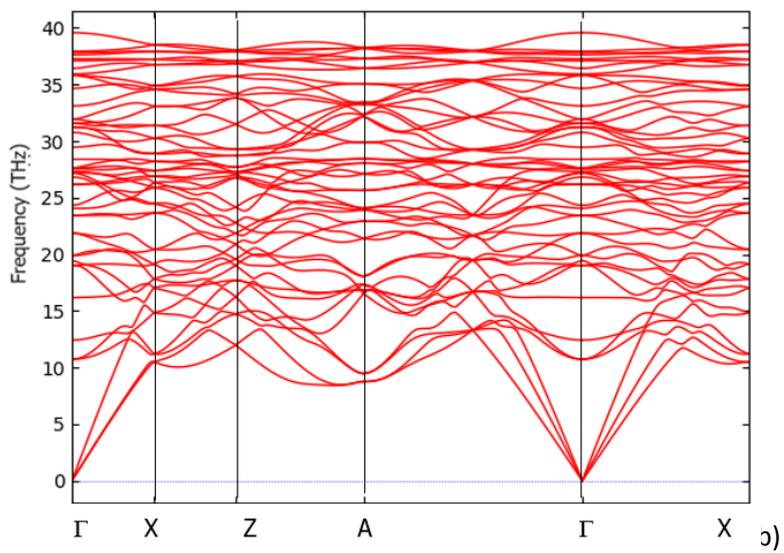

Figure 3. Phonons band structures: a) $C_{48}$, and b) $C_{32}$.

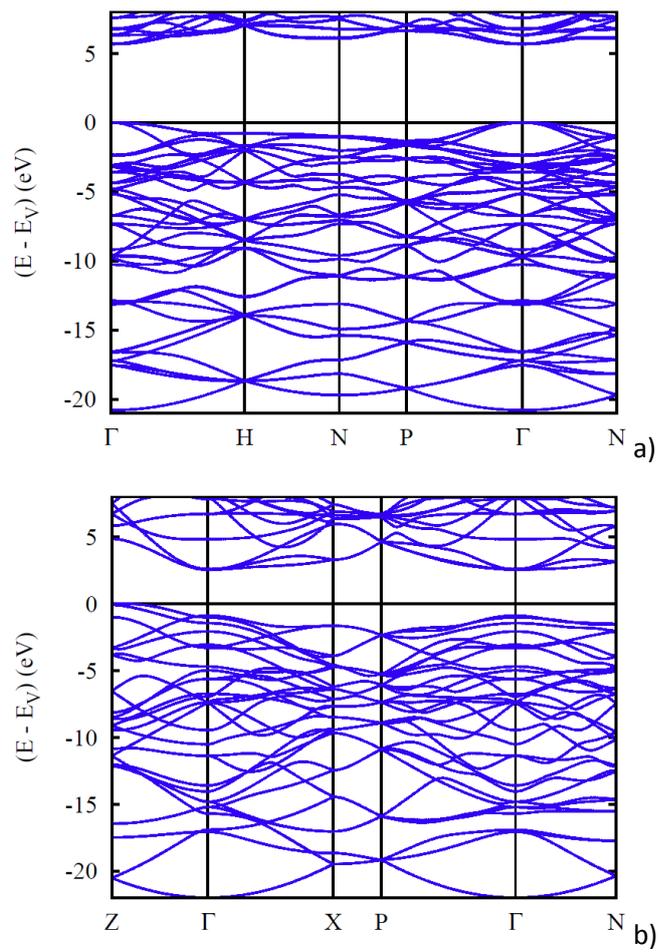

Figure 4. Electronic band structures: a) $C_{48}$, and b) $C_{32}$.